# Biological insights from RIL-seq in bacteria

Aviezer Silverman[1] and Sahar Melamed[1,2]

[1]Department of Microbiology and Molecular Genetics, Institute for Medical Research Israel-Canada, Faculty of Medicine, The Hebrew University of Jerusalem, Jerusalem, Israel

[2] Corresponding author: sahar.melamed@mail.huji.ac.il

**Running Head:** RIL-seq insights

## ABSTRACT

Bacteria reside in constantly changing environments and require rapid and precise adjustments of gene expression to ensure survival. Small regulatory RNAs (sRNAs) are a crucial element that bacteria utilize to achieve this. sRNAs are short RNA molecules that modulate gene expression usually through base-pairing interactions with target RNAs, primarily mRNAs. These interactions can lead to either negative outcomes such as mRNA degradation or translational repression or positive outcomes such as mRNA stabilization or translation enhancement. In recent years, high-throughput approaches such as RIL-seq (RNA interaction by ligation and sequencing) revolutionized the sRNA field by enabling the identification of sRNA targets on a global scale, unveiling intricate sRNA-RNA networks. In this review, we discuss the insights gained from investigating sRNA-RNA networks in well-studied bacterial species as well as in under-studied bacterial species. Having a complete understanding of sRNA-mediated regulation is critical for the development of new strategies for controlling bacterial growth and combating bacterial infections.

# 1. INTRODUCTION

Bacteria are known for responding and adjusting to environmental changes quickly and efficiently. This ability is facilitated by regulatory networks which tune the identity and quantity of the components needed for the bacterial response. Regulation in these networks is facilitated mainly at two levels, the transcriptional and the post-transcriptional level. Transcription factors that can act as activators or repressors control gene expression at the transcriptional level. The key players at the post-transcriptional level are regulatory small RNAs (sRNAs). These are mostly 50-400 nucleotide long RNA molecules that use sequence complementarity to bind to their RNA targets and regulate their stability and/or translation (reviewed in [1]). Most of the studied sRNAs regulate their targets *in trans* where the base-pairing is facilitated by an RNA chaperone. The most investigated RNA chaperone, primarily in Gram-negative bacteria, is the hexameric, ring-shaped Hfq protein [2]. While transcription factor networks have been studied thoroughly for decades [3], the mapping of sRNAs regulatory networks was lagging behind due to limited tools for finding sRNAs targets (reviewed in [4]). In recent years, several approaches were developed to provide a global map of sRNA-RNA interactions in bacteria (reviewed in [5]). Herein, we focus on the broadly used RIL-seq (RNA interaction by ligation and sequencing) methodology that integrates experimental and computational tools [6]. RIL-seq provides a comprehensive map of RNA-RNA interactions that take place on an RNA chaperone under a certain growth condition. In the experimental protocol, RNAs are *in vivo* crosslinked to the RNA chaperone (e.g., Hfq) by exposure to UV, the protein is immunoprecipitated with its bound RNAs, and neighboring RNA ends are then ligated. This gives rise to RNA chimeras that can be sequenced from both ends for the identification of RNA pairs. In the computational pipeline, reads are mapped to the genome and chimeric fragments are filtered for fragments that are statistically over-represented in the data to eliminate spurious interactions. RIL-seq was initially applied to *E. coli* K12 [7-9], but later it was used to reveal the sRNA interactome in multiple Gram-negative [10-13] and Gram-positive bacteria [14]. In the current review, we focus on the growth of knowledge in the regulation of gene expression in bacteria, gained by RIL-seq. We describe the expansion of sRNA networks, discuss a few examples of novel RNA sponges, elaborate on the advantage of using RIL-seq data for machine learning, highlight new properties of RNA-binding proteins, and revisit the roles sRNAs have in bacterial pathogenicity.

## 2. REVEALING INTRICATE RNA-RNA NETWORKS

sRNAs are a diverse group of gene expression regulators. While they act by different mechanisms [15], they also differ in the number of targets they have. In *E. coli* for example, RyhB is an iron metabolism regulator that has dozens of validated targets [16], while other sRNAs, like ChiX have only a handful of known targets [17]. Moreover, some targets can be regulated by several sRNAs, adding another level of complication, and establishing a connection between different pathways in bacteria. The sRNA-RNA networks drawn based on these interactions were useful but somewhat limited (e.g., resulting in 150 interactions in *E. coli* pre-RIL-seq; Table 1). RIL-seq technology serves as a tool for a significant expansion of these networks, uncovers sRNA-RNA networks in bacterial species where no network was known, and enables learning about the dynamics of the sRNA interactome. To date, RIL-seq dramatically expanded our knowledge in multiple Gram-negative bacteria and in one Gram-positive bacteria (Table 1). Valuable online platforms to visualize the RNA-RNA networks and RIL-seq data were generated and are summarized in Table 2.

The extensive networks drawn by RIL-seq opened an uncharted world of sRNA-mediated regulation. This includes the identification of novel sRNAs from previously unimagined regions, such as internal to the coding sequence (CDS) and from other regions, such as the 3' UTR or 5' UTR, expansion of the target sets of known sRNAs, discovery of new sRNA-sRNA interactions, interactions involving tRNAs, and mRNA-mRNA interactions. While a portion of these interactions represent canonical sRNA mechanisms of action (reviewed in [15]), some require more thorough deciphering. For others, we do not have sufficient knowledge to understand their meaning. A realization resulting from these valuable datasets is that sRNA-mediated regulation is as abundant as transcription factor-mediated regulation, emphasizing the complexity of regulation of gene expression in bacteria. In different bacterial species, further insights can be gained. Enteropathogenic *E. coli* (EPEC) can be grown in conditions that resemble those in the host environment and induce the expression of virulence genes (activating conditions) or in conditions that do not resemble the host environment (non-activating conditions)[18]. Pearl Mizrahi et al. carried out RIL-seq in non-activating and in activating conditions. Under non-activating conditions, ~16% of the RIL-seq chimeras involved genes from accessory regions (e.g., regions that encode pathogenicity traits) whereas under activating conditions this percentage increased to ~44%, implying a more significant involvement of sRNAs in EPEC pathogenicity than previously thought [12]. In *Salmonella*, a total of ~1240 significant chimers were found suggesting 74 sRNAs, 25 of which were previously characterized as Hfq-dependent sRNAs, multiplying by three the number of Hfq-dependent sRNAs [11]. RIL-seq done in *Pseudomonas aeruginosa* uncovered that a single sRNA called PhrS dominates the RNA-RNA interaction network by pairing with ~800 targets [13]. In *E. coli*, the application of RIL-seq to two different RNA chaperones revealed that the RNA chaperones compete for the same RNA pairs, playing opposite roles [8]. As more RNA-RNA networks are expected to be published, computational analysis followed by experimental ones will be able to give us new insights about the evolution of sRNAs and their targets.

## 3. IDENTIFICATION OF RNA SPONGES

Traditionally, sRNA functions and their ability to regulate gene expression were studied through the mRNA targets to which they were binding. However, an increasing number of independent studies in recent years documented the binding of two sRNAs, where one regulates the activity of the other, a mechanism that we now call "sponging." Since the focus has been on documenting effects of a sRNA on its target, a target that was not affected by the sRNA was ignored and therefore RNA sponges have been overlooked. The concept of RNA sponges was first described for microRNAs (miRNAs) by Albert et al. in 2007 and was called "target mimicry" [19]. In this study, they showed that by engineering a synthetic RNA that carried repeats of specific sequences, it could base pair with other known miRNAs and compete with their natural targets, subsequently causing the removal of this miRNA from the regulatory pool and depleting its action. The first RNA sponge to be identified in bacteria was an intergenic region between *chbB-chbC*, two genes that are part of the chitobiose operon. It was shown to sponge a sRNA called ChiX, allowing for the expression of ChiX targets that were previously repressed by ChiX [20, 21]. This discovery was followed by the identification of two other RNA sponges in *E. coli* and *Salmonella*, SroC, sponging the global regulator GcvB [22], and 3'ETS$^{leuZ}$, an RNA fragment processed from a tRNA precursor, that regulates the sigma E-dependent, iron homeostasis related sRNA RyhB [23]. The mechanism by which RNA sponges act involves base pairing with a sRNA, depleting its availability in the cell and possibly causing its degradation.

The development of various high-throughput RNA-sequencing-based approaches allows unbiased identification of RNA sponges. Herein, we discuss the identification of RNA sponges by RIL-seq (Summarized in Figure 1) while deliberation on sponges found by other approaches can be found elsewhere [24]. RIL-seq analysis typically yields between a few hundred to thousands of statistically significant RNA pairs. Of which, a significant subset are sRNA-sRNA interactions, providing a large pool of potential RNA sponges.

Matera et.al. applied the RIL-seq to *Salmonella enterica* during early stationary phase and they identified a new potential target of the classic porin repressor MicF [25]. As expected, MicF's primary target, the outer membrane porin *ompF*, was found with MicF in many chimers. However, the analysis recovered nearly twice as many MicF chimeras with the major inner membrane transporter mRNA, *oppA*. This observation was surprising because previous work on MicF targets did not find any change in the translation of the *oppABCDF* operon [26, 27]. The base pairing between MicF and *oppA* was actually taking place upstream of the start codon, in a region from which a separate transcript was generated, and they called this transcript OppX. Further research classified this 5' UTR as a major sRNA sponge binding MicF and depleting its availability in the cell without causing its degradation, hence acting as a "soaking" sponge. 5' UTR sRNAs can be easily missed due to limited transcriptome annotation, whereas RIL-seq analysis can point to potential 5' UTR-originated sRNAs. The identification of OppX adds another level of complexity to the regulatory network of MicF and helps to better understand this regulatory circuit. In another study, Huber et al. applied RIL-seq to *Vibrio cholerae* and found 81 sRNA-sRNA interactions [10]. Out of all those sRNA pairs, a novel sRNA drew the authors' attention as the data suggested that it base pairs exclusively with the four Qrr sRNAs known to modulate the quorum sensing pathway in *V. cholera*. The sRNA was termed QrrX and the characterization of its interaction with the Qrrs revealed that it functions as an RNA sponge, affecting the levels and activity of the Qrr sRNAs. Specificity of RNA sponges was also observed in the case of PspH in *E.coli* [9]. PspH is a 3' UTR-derived sRNA, interacting solely with Spot 42 sRNA, a key player in the regulation of carbon transport and metabolism. Overexpression of PspH led to an almost 5-fold reduction in Spot 42 levels and to increased levels of several Spot 42 targets. RbsZ is another RNA sponge in *E. coli* identified by RIL-seq and its characterization uncovered an additional level of complexity to sRNA-mediated regulation [8]. In this case, Melamed et al. demonstrated that RIL-seq could be applied to other RNA chaperones than Hfq and they applied it to both Hfq and the understudied ProQ. A comparison of the RNA pairs found on Hfq and ProQ revealed that about a third of the RNA pairs captured on ProQ were also found on Hfq, suggesting overlapping and/or competing roles for the two RNA-binding proteins. One of these overlapping RNA pairs was between RybB, a sRNA known to be active under cell envelope stress, and a novel sRNA, RbsZ. RbsZ corresponds to the 3′ UTR of *rbsB*, which is part of the ribose catabolism operon. Characterization of the RybB–RbsZ interaction revealed a fascinating interplay between two RNA-binding proteins. It was found that RbsZ downregulates RybB by base-pairing, leading to RybB degradation, and relieving the RybB-dependent regulation on its targets. RbsZ-dependent regulation was mediated by Hfq and blocked by ProQ, resulting in a continuous competition between the two RNA-binding proteins on this RNA pair.

The increasing number of RNA sponges found in recent years enhances our understanding of sRNA mechanisms of action. While previously the regulation of sRNA activity was thought to be achieved mainly by controlling its expression, it is now acknowledged that post-transcriptional regulation by RNA sponges is just as important. Most of the studied RNA sponges to date share two main traits in the RIL-seq data: 1. An RNA is found exclusively in chimeras with another sRNA (e.g., does not have any other RNA partners), and 2. It is ranked high in that sRNA target set based on the number of chimeras. The examples above were taken from the studies in which RIL-seq was performed. Later studies used the published RIL-seq datasets to uncover additional RNA sponges [28], demonstrating how RIL-seq is an extremely valuable tool for identifying RNA sponges in regions that were not previously thought to encode sRNAs.

**4. IDENTIFICATION OF NOVEL SRNAS USING MACHINE LEARNING**

Initially, bacterial-encoded sRNAs such as Spot 42 [29, 30], MicF [31], and OxyS [32], were found sporadically and thought to be encoded only in intergenic regions. Today, it is well documented that sRNAs can be encoded from diverse genomic loci including portions of mRNAs and tRNAs (reviewed in [33]), making it harder to detect them systematically. The development of high-throughput techniques, like the microarray at the beginning [34] and RNA sequencing a decade later (reviewed in [35]), allowed for the identification of an increasing number of sRNAs. The development of RNA-seq-based techniques that define RNA boundaries, such as differential RNA-seq [36] or Term-seq [28, 37], gave another boost to the identification of sRNAs. However, while all these methods could point on the existence of sRNAs, they could not provide any data on the functionality of these short RNA fragments expressed from the genome.

The RIL-seq data withholds valuable information for the interacting RNAs that can be interpreted to better understand the function of the interacting RNAs (e.g., target or regulatory RNA). Bar et al. have nicely developed a machine learning algorithm to identify sRNAs from RIL-seq data based on several characteristics of the RNAs in the data [38]. The authors used 26 to 29 known sRNAs as a training dataset to detect significant features of sRNAs and applied the machine learning algorithm on the remaining ~1000 RNAs found in the *E. coli* Hfq RIL-seq data, in each growth condition [9]. By doing so, they were able to identify in high probability 69 sRNAs, 28 of which served as the training set, 18 were experimentally confirmed by the authors or by other groups recently, and 23 were defined as promising sRNA candidates (Figure 2). These findings contribute to the scientific community in multiple ways. First, this study demonstrates that there are specific characteristics of sRNAs in the RIL-seq data that can be used to predict novel sRNAs. This will be extremely useful in less-studied bacteria where only a handful of sRNAs are known, and the characteristics of these sRNAs in the RIL-seq data can be used to train the algorithm for the prediction of novel sRNAs. Second, RIL-seq analysis can be overwhelming due to the amount of data it generates, especially when most of the findings are yet to be understood and are not well documented. The ability to study and analyze all of this information experimentally is limited and combining computational tools with experimental ones can be a game changer for successful research.

## 5. EXPANSION OF RNA-BINDING PROTEINS PROPERTIES

RNA molecules are naturally unstable and thus need to be synthesized in large numbers, form stable secondary structures, or be protected by other proteins. For example, the bacterial mRNA is transcribed and simultaneously translated into a protein. This way, the mRNA is protected from degradation by RNases and once the translation is complete, the molecule is most likely to be degraded and recycled [39]. In contrast, sRNAs are not translated and therefore may be exposed to RNases and quickly turned over. RNA-binding proteins bind to specific RNA molecules at specific sites based on sequence or structural recognition (reviewed in [40]). These proteins are often referred to as chaperones because they mediate the function of the RNAs. As mentioned earlier, Hfq binds both the sRNA and its target mRNA, and facilitates the sRNA-dependent regulation, is the most studied one [2, 41]. Yet, the complete role Hfq plays in sRNA-mediated regulation is still not fully understood. The extensive data achieved by RIL-seq can be used to learn more about the characteristics of RNA chaperones. For example, the main region by which a sRNA binds to Hfq is through its GC-rich region followed by a poly-U tail, which represents the Rho-independent transcriptional terminators. This specificity implies that the 3' end of the sRNA is buried in the Hfq hexamer and is less available for the ligation step in the RIL-seq protocol. This resulted in a strong enrichment for sRNAs located on the 3' end of the RIL-seq chimeras recapitulating what was already known. Faigenbaum-Romm et al. used *E. coli* Hfq RIL-seq data to gain new insights about the hierarchy in the chaperone occupancy of sRNA targets [7]. The authors estimated the binding capabilities of two RNA molecules by calculating their free energy using two tools (RNA duplex and RNAup), and the amount of Hfq bound to a single RNA from the read counts (single and chimeric) from RIL-seq experiments. While they observed a weak correlation between the free energy of a bound sRNA to its target and the number of chimeric fragments, they did find a strong correlation between the sRNA-target interaction frequency and the Hfq occupancy of the target in a specific condition. This suggests that the ability of one target to interact with the Hfq protein has a large impact on its level of regulation by its corresponding sRNA. This competition among targets explains the difference between affected and unaffected RNAs by the sRNA Hfq-mediated regulation. They found that sRNAs that bind many Hfq molecules have a more extensive impact on gene regulation as the amount of Hfq in the cell is closely regulated and tends to be maintained within a certain concentration range. Thus, sRNAs with better binding affinity for Hfq are more likely to have a more substantial effect on the cell's physiology. Gaining a better understanding of how RNA chaperones facilitate the regulation of gene expression will also help to develop new ways to manipulate them.

While RIL-seq was initially applied to Hfq, it can be applied to other RNA-binding proteins, as described earlier in this review [8]. The fact that ProQ blocked the Hfq-mediated RbsZ-dependent regulation intrigued the authors to hypothesize that the protective ability of ProQ might be broader than just affecting one RNA pair. To test this idea, ProQ was briefly overexpressed and the effect on the *E. coli* transcriptome was studied. The authors found that ProQ was protected many transcripts. This finding expanded the role of ProQ and suggested that it has a global role in protecting and/or stabilizing RNA transcripts. The interplay between the *E. coli* Hfq and ProQ opens the door for other types of relationships between RNA-binding proteins to be studied in the future (reviewed in [42]).

In conclusion, RIL-seq enables an expansion of our knowledge regarding proteins that take part in the regulation of gene expression (Figure 3), and it can be applied to other RNA-binding proteins of interest such as CsrA [43] or others that were recently found [44]. It suggests that sRNA-dependent regulation is more complex than previously thought and that RNA-binding proteins have multiple means to fine tune sRNA-dependent regulation.

## 6. UNCOVERING SRNAS INVOLVED IN PATHOGENICITY

Bacterial pathogens play a crucial role in many environments, including the human body. Bacterial species often acquire pathogenicity genes through mutational events, horizontal gene transfer, and bacteriophage infections [45]. Performing RIL-seq in pathogenic bacteria provides exclusive opportunities to understand better how sRNAs couple the core genes with virulence traits encoded by these additional genes at the post-transcriptional level (Figure 4). To date, RIL-seq analysis was performed in several common pathogens, such as EPEC [12], *Salmonella* [11], *Clostridioides difficile* [14], and *V. Cholera* [10].

EPEC has a similar genome to that of *E. coli* K-12, genes that are identical are called the core genome, whereas additional genes that are pathogenesis associated are called "accessory genes" and cluster in pathogenic islands. Pearl Mizrahi et al. applied RIL-seq to EPEC and cross referenced the RNA-RNA interactions with non-pathogenic *E. coli*. Many interactions were shared between the strains, but a subset of the chimeras involved the accessory genome. They hypothesized that these interactions are important for the pathogenicity of EPEC. In this study, EPEC was grown in two distinct environments, one causing the activation of pathogenic traits, such as quorum sensing and aggregation, and the other non-activating. RIL-seq revealed a significant difference in the RNA interactome between these two growth conditions. Interestingly, they noticed that the abundance of several sRNAs, like Spot 42, RyhB, MgrR, and CpxQ, under the non-activated condition was higher and that the core sRNA network changed under activating conditions, implying sRNAs regulate EPEC pathogenicity. The authors focused on MgrR, a sRNA that had more interactions in the activating state than the inactivated state and had chimeras with accessory genes. The top target of MgrR was *cesT*, encoding the major virulence-associated chaperone. MgrR adjusted the level of EPEC cytotoxicity via regulation of *cesT* expression and by enhancement of the transcription of all the LEE (locus of enterocyte effacement) genes, including that of *cesT*, through the regulation of another target.

*Salmonella* is one of the leading bacterial foodborne causes of death in low and middle-income countries [46-48]. As Gram-negative bacteria, they have a complicated system of inner membrane and outer membrane proteins that must balance membrane integrity and nutrient uptake. Matera et al. studied the sRNA interactome in *Salmonella* and provided important insights on the fine regulation of membrane permeability controlled by sRNAs [11]. As discussed earlier in this review, OppX is a novel sRNA that acts as an RNA sponge and regulates MicF, a well-studied prototypical porin repressor. In this work, the bacteria were grown in conditions that express the *Salmonella* pathogenicity island 1 (SPI-1) virulence genes, which encodes many virulence proteins and a type III secretion system (T3SS) for their secretion. Interestingly, while previous target searches for the InvR sRNA encoded by SPI-1 excluded effects of this sRNA on SPI-1 virulence gene expression [49], RIL-seq data reveal many chimeras between InvR and SPI-1 encoded mRNAs (Table S2 in [11]). Further studies will required to determine the nature of these interesting findings and their significance for *Salmonella* pathogenicity.

Quorum sensing is the ability of a bacterium to sense a signal from neighboring cells in the population when they reach a critical concentration and synchronously control processes that are only efficient above this cell concentration. This ability allows pathogenic bacteria to minimize host immune responses by delaying the production of virulence factors until sufficient bacteria are present and prepared to overcome the host defense mechanisms [50]. One of the model bacterial species to study quorum sensing is *Vibrio*. Recently, RIL-seq analysis was applied to *V. cholera* [10], identifying hundreds of sRNA-mRNA interactions and RNA duplexes. One novel sRNA had an interesting interactome, in that it exclusively interacted with all four quorum regulator RNAs (Qrr1-4). As described above in this review, further experiments confirmed the hypothesis that this sRNA, termed QrrX, acts as an RNA sponge and adds another layer of regulation to the quorum sensing pathway in *Vibrio*. In *P. aeruginosa*, the PhrS sRNA regulates quorum sensing by upregulating the transcription factor MvfR required for the synthesis of quorum sensing signaling molecules [51]. A RIL-seq study revealed that PhrS negatively regulates the *antR* transcript that encodes AntR, a regulator that directs a precursor of quorum sensing signaling molecules called anthranilate towards the TCA cycle [13]. Thus, PhrS upregulates the synthesis of quorum sensing signaling molecules through a two-tiered mechanism, the upregulation of *mvfR* and downregulation of *antR*.

Gram-positive pathogens such as *C. difficile* pose a significant challenge to healthcare systems as they form antibiotic-resistant endospores that lead to relapsing and recurrent infections. Understanding the molecular mechanisms and the environmental signals that control spore generation is essential for developing efficient treatments. Fuchs et al. applied RIL-seq, for the first time in Gram-positive bacteria, to *C. difficile* under sporulating conditions [14]. The authors identified sRNA-mediated post-transcriptional regulation of *spo0A*, the master regulator of sporulation, as a new mechanism contributing to sporulation initiation in *C. difficile*. They found two novel sRNAs, SpoY and SpoX, that bind to *spo0A* mRNA and results in inhibition or activation of translation, respectively.

The emergence of antibiotic-resistant bacteria is an increasing global concern [52]. It is suggested that by the year of 2050, as many as ten million people could die each year due to resistant microbes [53]. Deciphering the mechanisms of bacterial pathogenicity is crucial for developing novel effective therapeutics. As we discussed above, regulatory RNAs in pathogenic bacteria participate in such mechanisms and a further understanding of their mechanism of action can help develop novel antimicrobial agents [54].

## 7. PERSPECTIVES

While our knowledge of sRNA regulatory networks and modes of action is growing by the day, there is still much left to learn. The growing use of RIL-seq and other RNA-RNA identification methods such as CLASH [55] or Hi-GRIL-seq [56] provides vast amounts of data that can be mined to deepen our understanding of sRNA-mediated regulation. Current and future studies will take advantage of this knowledge to decipher specific mechanisms by which sRNAs act and how they integrate with other regulatory pathways, constituting multiplayer regulation of gene expression. The gained knowledge can be used for development of therapeutics targeting sRNAs or using them for targeting other genes in bacterial pathogens. Furthermore, integration of RNA sequencing-based methods with other multi-omics data, such as genome sequencing and metabolomics, will lead to a more comprehensive understanding of the role sRNAs play in bacteria.

**DISCLOSURE STATEMENT**

The authors are unaware of any affiliations, memberships, funding, or financial holdings that might be perceived as affecting the objectivity of this review.

**ACKNOWLEDGMENTS**

We appreciate the useful comments from Philip Adams. We are grateful to the Melamed lab members for helpful discussions and comments on the manuscript. The study was supported by the Israel Science Foundation (grants 826/22 and 2859/22). Figures 1, 3 and 4 were created with BioRender.com.

**FIGURE LEGENDS**

Figure 1

RNA sponges regulate the activity of sRNAs, affecting diverse pathways in a range of bacterial species. QrrX base-pairs with Qrr1-4 sRNAs in *V. cholerae*, leading to their degradation and impairing quorum sensing regulation. In *E. coli*, RbsZ binding to RybB leads to its degradation, enhancing the synthesis of outer membrane porins. Downregulation of Spot 42 by PspH enables the expression of transport and secondary carbon sources utilization genes. The OppX sRNA positively affects the synthesis of the key porin OmpF in *Salmonella* by sponging MicF. The schematic bacterial cell drawing does not represent a specific species.

Figure 2
Visualization of the results of machine learning algorithm based on RIL-seq data [38]. The graph shows the number of known sRNAs that were used as a training set for the algorithm (blue), predicted sRNAs via machine learning that were confirmed by the authors and others (green), and predicted sRNAs via machine learning that are yet to be experimentally verified (gray).

Figure 3

The binding properties between RNA-binding proteins and the RNAs they bind govern the outcome of sRNA-mediated regulation. The binding efficiency of an mRNA target to Hfq (green) affects its fate. ProQ (red) can bind sRNA-mRNA pairs and thus compete with Hfq.

Figure 4

sRNAs regulate bacterial pathogenicity in multiple species. The *Salmonella* SPI-1-encoded sRNA InvR was found with many SPI-1 mRNA targets in RIL-seq data. Sporulation in *C. difficile* is regulated by two sRNAs, SpoY and SpoX, with opposite outcomes. PhrS promotes the synthesis of quorum sensing signaling molecules in *P. aeruginosa* by regulating two transcription factors, MvfR and AntR. The network of QrrX with the four Qrr sRNAs controls virulence and biofilm formation, among other pathways in *V. cholerae*. EPEC cytotoxicity is regulated by the core genome-encoded sRNA MgrR.

**TABLES**

| Bacterial species | pre RIL-seq estimated sRNA network | RIL-seq network | | | | Reference |
|---|---|---|---|---|---|---|
| | | Growth conditions | RNA chaperone | Number of interactions | Unique interactions | |
| *E. coli* | *~150* | Log phase | Hfq | 1027 | 3138 | [9] |
| | | Stationary phase | Hfq | 1844 | | |
| | | Iron limitation | Hfq | 1947 | | |
| | | Rich media | Hfq | 2069 | 2413 | [8] |
| | | Minimal media | Hfq | 1109 | | |
| | | Rich media | ProQ | 322 | 320 | |
| | | Minimal media | ProQ | 33 | | |
| *EPEC* | *~155* | Activating | Hfq | 810 | 1555 | [12] |
| | | Non-activating | Hfq | 1050 | | |
| *P. aeruginosa* | *~300* | Exponential phase | Hfq | 997 | 1446 | [13] |
| | | Stationary phase | Hfq | 702 | | |
| *Salmonella* | *<100* | Early stationary 1 | Hfq | 883 | 1240 | [11] |
| | | Early stationary 2 | Hfq | 891 | | |
| *V. cholerae* | *<100* | Low cell density | Hfq | 3193 | 4274 | [10] |
| | | High cell density | Hfq | 1860 | | |
| *C. difficile* | *<5* | sporulation conditions | Hfq | 1569 | 1186 | [14] |

**Table 1**: A summary table of the number of RNA-RNA interactions identified by RIL-seq in different studies. "Unique interactions" represents the number of unique interactions that were found in each study. Each pair was counted only once except for cases where the same pair appeared in a different order in the chimera, and then it was counted twice.

| Platform | Bacterial species | Description | Data from | Link |
|---|---|---|---|---|
| UCSC genome browser | *E. coli* | Hfq RIL-seq single fragments data | [9] | Single fragments |
| | | Hfq RIL-seq chimeric fragments data | | Chimeric fragments |
| | | Hfq and ProQ RIL-seq data | [8] | Hfq and ProQ |
| | | Hfq and ProQ RIL-seq data in *hfq* and *proQ* deletion backgrounds | | *hfq* and *proQ* deletion backgrounds |
| | *P. aeruginosa* | Hfq RIL-seq data in exponential growth phase | [13] | exponential phase data |
| | | Hfq RIL-seq data in stationary growth phase | | stationary phase data |
| RILseqDB | *E. coli* and EPEC | A user-friendly interface to query and view the results of RIL-seq experiments. Interactions of specific RNA can be queried | [9, 12] | http://rilseqdb.cs.huji.ac.il/ |
| RNA-RNA interactome browser | *Salmonella* | The browser allows for an interactive search on sRNAs and coding genes, thereby enabling an intuitive visualization of sRNA regulons, and offering users the option to upload and visualize additional RIL-seq datasets | [11] | http://resources.helmholtz-hiri.de/rilseqset/ |
| RILSeqExplorer V. cholerae | *V. cholerae* | A dynamic and searchable web interface that provides a network view of RIL-seq interactions | [10] | http://rnaseqtools.vmguest.uni-jena.de/ |
| RNA-RNA interactome browser | *C. difficile* | *C. difficile* RIL-seq network | [14] | https://resources.helmholtz-hiri.de/rilseqcd/ |
| sInterBase | *E. coli* | Platform for mining sRNA-mRNA interactions in *E. coli* | [57] | https://sinterbase.cs.bgu.ac.il/ |

**Table 2**: A summary of online platforms for visualization and analysis of RIL-seq data.

Figure 1

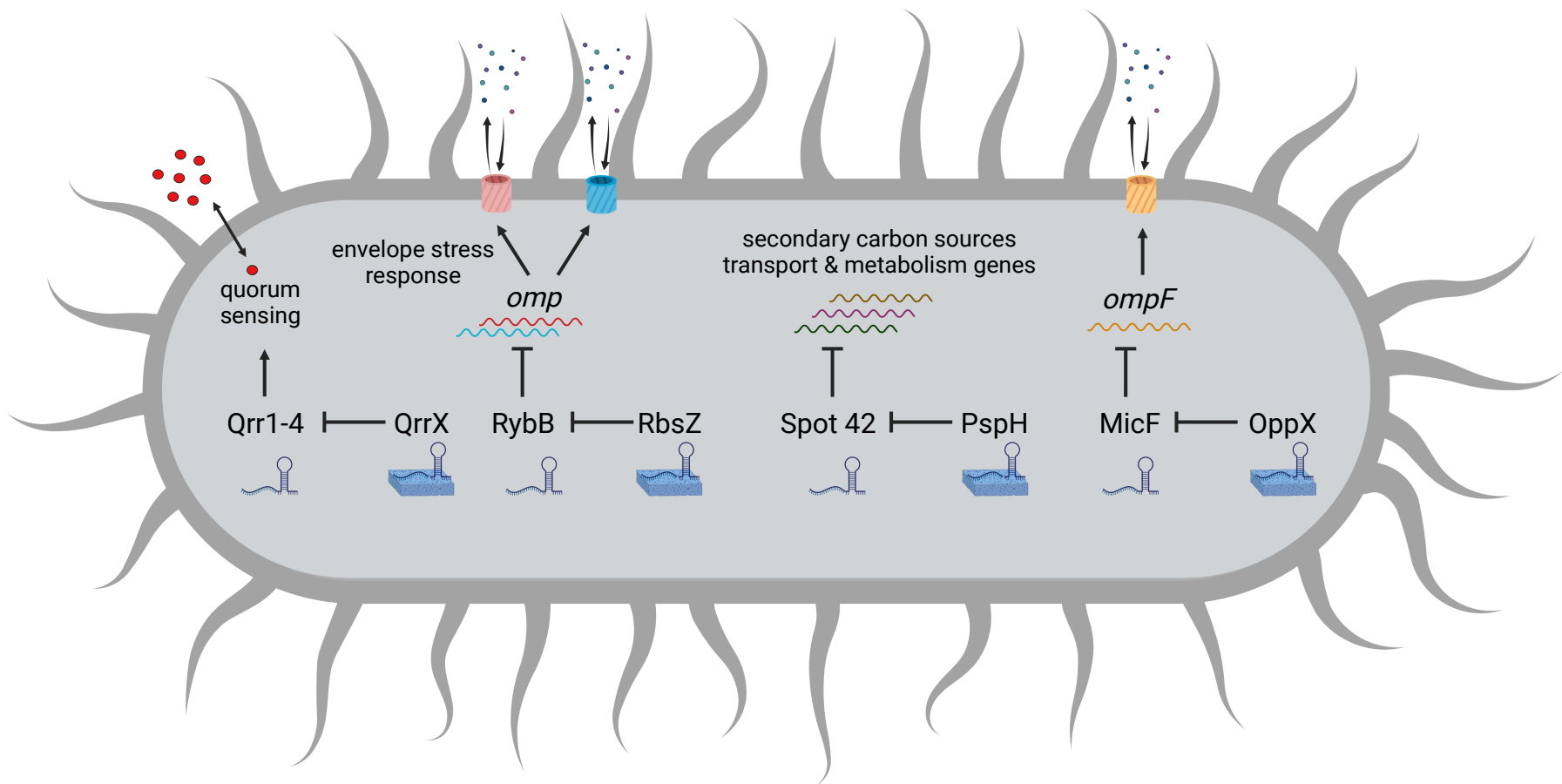

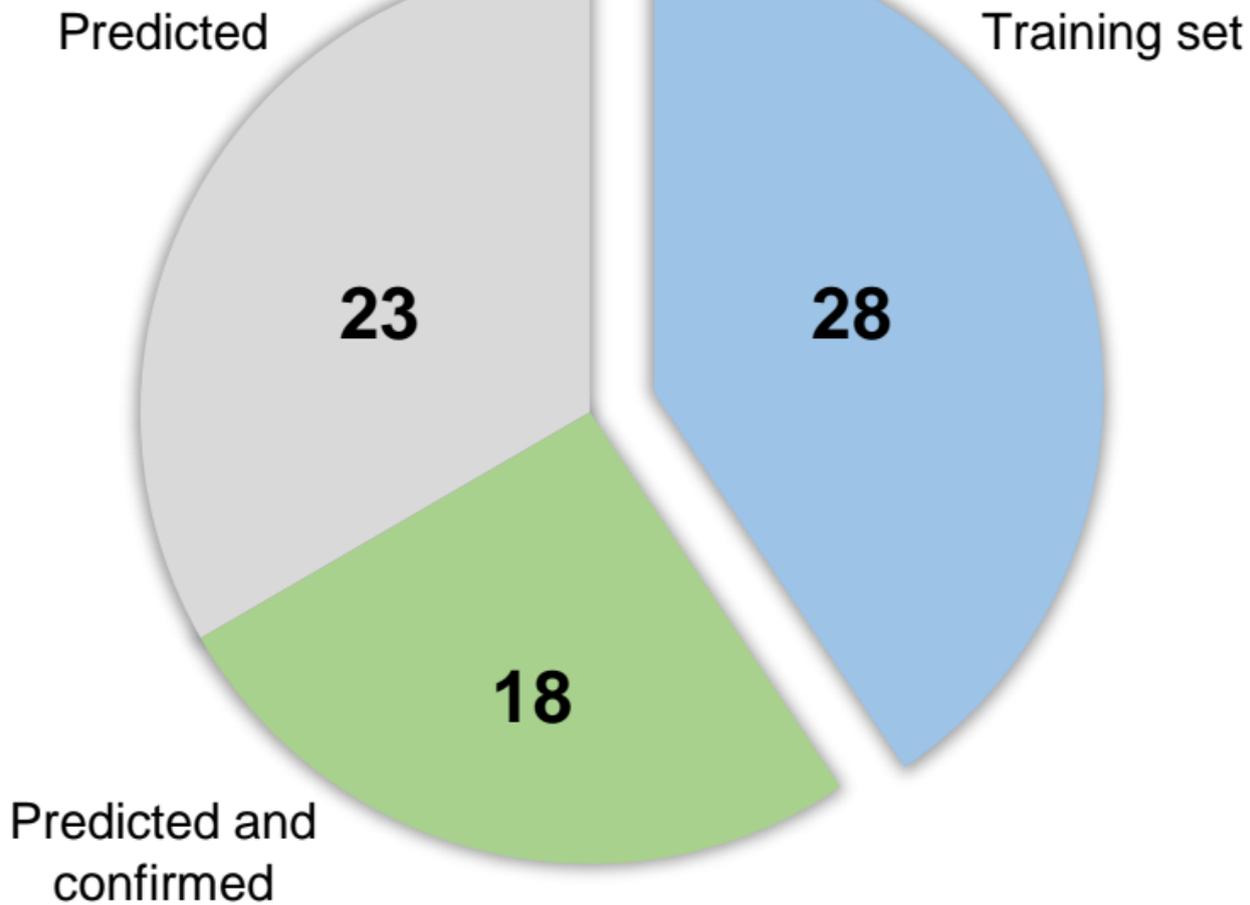

Figure 2

Figure 3

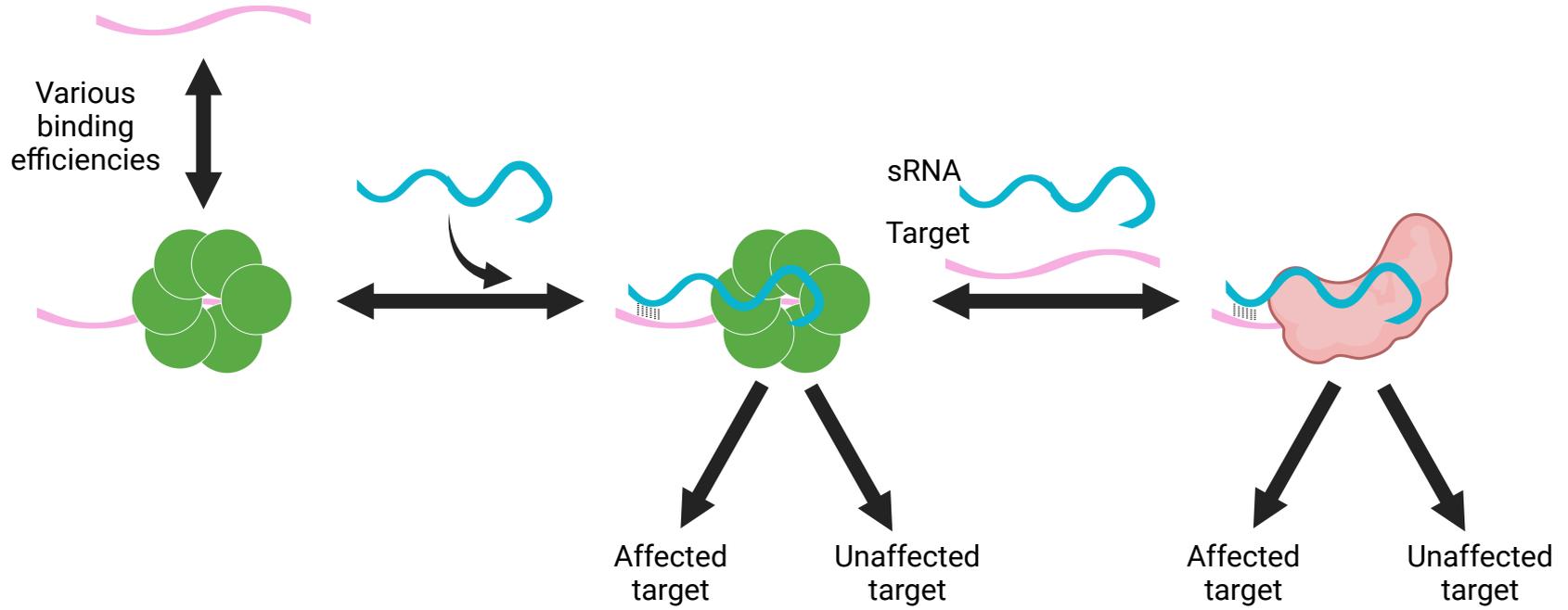

Figure 4

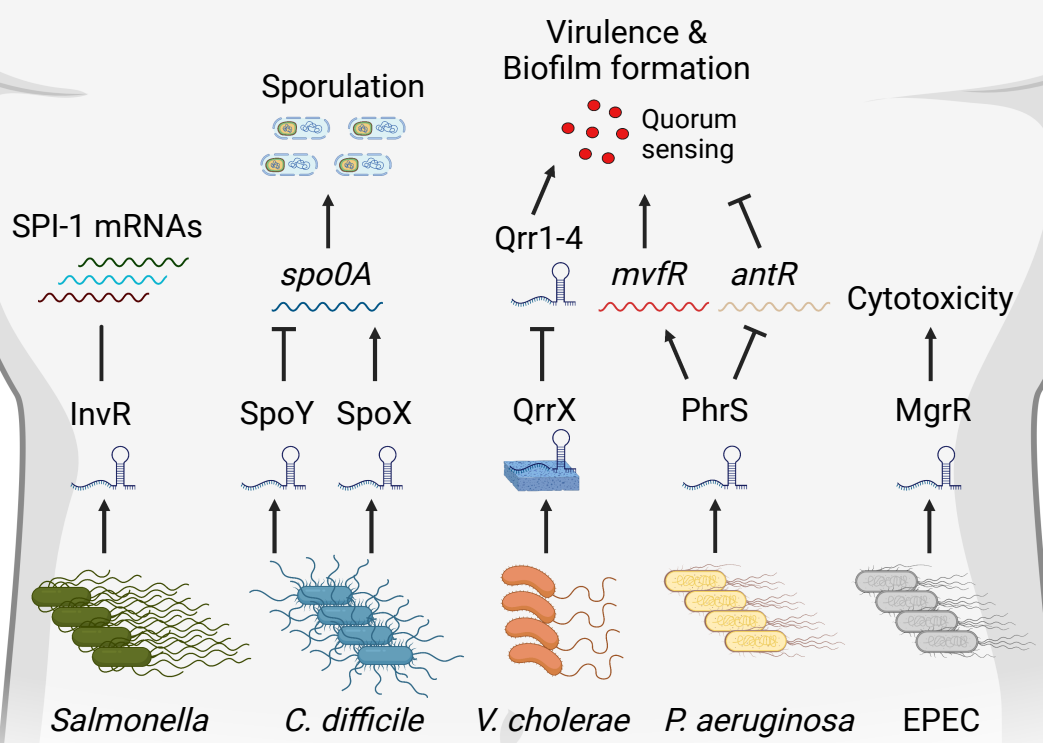